\documentclass[showpacs,preprint,aps]{revtex4}
\begin{document}
\title{
Ab-initio study of model guanine assemblies:
the role of $\pi$-$\pi$ coupling and band transport
}
\author{
Rosa Di Felice, Arrigo Calzolari, and Elisa Molinari
}
\affiliation{
Istituto Nazionale di Fisica della Materia (INFM) and Dipartimento di Fisica, 
Universit\`a di Modena e Reggio Emilia, Via Campi 213/A, 41100 Modena, Italy
}
\author{
Anna Garbesi
}
\affiliation{
CNR ISOF, Area della Ricerca, Via P. Gobetti 101, 40129 Bologna, Italy
}
\begin{abstract}
Several assemblies of guanine molecules 
are investigated by means of first-principle calculations. 
Such structures include stacked and hydrogen-bonded 
dimers, as well as vertical columns and planar ribbons, respectively, 
obtained by periodically replicating the dimers. 
Our results are in good agreement with experimental 
data for isolated molecules, isolated dimers, and periodic ribbons. 
For stacked dimers and columns, the 
stability is affected by the relative charge distribution 
of the $\pi$ orbitals in adjacent guanine molecules. 
$\pi$-$\pi$ coupling in some stacked columns induces 
dispersive energy bands, while no dispersion 
is identified in the planar ribbons along the 
connections of hydrogen bonds. 
The implications for different materials comprised of 
guanine aggregates are discussed. 
The bandstructure of dispersive configurations may 
justify a contribution of band transport (Bloch type) 
in the conduction mechanism of 
deoxyguanosine fibres, 
while in DNA-like configurations band transport should be negligible. 
\end{abstract}
\maketitle
%
\section{Introduction.}
\protect\label{sec-introduction}
The twofold issue of scaling down the electronic devices and 
of realizing a high circuit integration density on a single chip 
has recently seen the rise of the field of molecular electronics, which 
consists of the use of molecules to realize electrically conducting 
structures \cite{rev-aviramratner,rev-gimzewski}.
Because of their sequence-specific recognition properties, DNA 
molecules are attracting attention for the construction of 
nanometer scale devices, where they might be used, by virtue of 
their self-assembling capabilities, to wire the electronic materials 
in a programmable way \cite{braun-nat98}. 
This research path has led recently to a set of controlled 
experiments for the direct measurement of the d.c. conductivity. 

Using interdigital electrodes, anisotropic conductivity 
was found in an aligned DNA cast film: At room temperature, 
a large ohmic current, linearly increasing with the applied voltage, 
was measured 
\cite{okahata-jacs98}. 
Ohmic behavior and high conductivity were found, also, 
for a a 600 nm long rope made f a few $\lambda$-DNA molecules 
\cite{fink-nat99}. Instead, nonlinear current/voltage curves, 
exhibiting a voltage gap at low applied voltage, 
were measured through a single 10 nm long poly(dG)/poly(dC) 
DNA molecule trapped between two metal nanoelectrodes 
\cite{porath-nat00}. 
Large currents were observed, in air and in vacuum, 
both at ambient temperature and at 4 K. The authors suggested that 
the observed electron transport is best explained by a 
semiconductor-like band model where the electronic 
states are delocalized over the entire length of the 
base pair stack 
\cite{porath-nat00}. 

The guanine (G) base is particularly interesting in view 
of obtaining conductive molecular aggregates, because of its 
low ionization potential, which suggests its viability 
to mediate charge motion along a sequence of bases 
\cite{meggers,jortner98}. 
This peculiar property has opened the way to the measurement of the 
the electrical conductivity of G aggregates. 
Interestingly, a metallic nanogate 
filled with a dried solution of a lipophilic derivative 
of 2'-deoxyguanosine \cite{apl,prl}, displayed a current/voltage behavior 
\cite{note-counterions} 
similar to that of the semiconducting 
poly(G)/poly(C) sample 
\cite{porath-nat00}. 
Previous investigations 
had shown that, in organic solvents, this molecule undergoes 
extensive self-assembly, mediated by H-bonding among the guanine bases, 
to give ribbon-like aggregates that, upon drying, form fiber 
structures within which the guanine cores of the ribbons 
lie on parallel planes at a distance of 
about 3.4 \AA~\cite{gottarelli98,gottarelli00}.
While a direct comparison among the findings 
for the deoxyguanosine materials and for DNA cannot be made, 
because of the different overall experimental 
settings and chemical nature of the molecules, one cannot avoid 
to notice the similar and peculiar current/voltage 
characteristics of the nanogates interconnected by molecules 
featuring self-assembled \cite{apl,prl} or inherent \cite{porath-nat00} 
guanine stacks. 

In the following, we have chosen to perform ab-initio 
calculations of the structural and ground state electronic properties 
of extended model structures whose building block is the 
G base alone. Of course, we are well aware that 
these model structures are only partially related to the 
structure of a poly(G)/poly(C) duplex \cite{porath-nat00}, where each guanine 
is H-bonded to a cytosine in the opposite strand. 
However, self-assembled G structures are among the simplest base 
aggregates characterized experimentally 
\cite{prl,gottarelli98}, thus allowing accurate 
theoretical calculations as well as 
comparison with experimental data. 
Additionally, our choice 
is motivated by the role played by guanine, 
the DNA base with the lowest ionization potential, in the 
mechanism of charge transport 
\cite{bixon-hhop99,schuster-polaron,GGenhancedtransport}. 

The main question that we address here is whether the electronic 
properties of extended G-based structures can account for a band-like 
mechanism for charge transport. Previous theoretical studies for 
DNA bases and for their assemblies can 
be found in the existing literature 
\cite{sponer1,sponer2,sponer3,sugiyama-saito,machado,parrinello}. 
Both MP2/Hartree-Fock and density-functional-based
calculations were performed
in the past with localized basis sets (GAUSSIAN) to treat
isolated guanines and small clusters of guanines
\cite{sponer1}. Here, we focus on density-functional-based
calculations using a plane wave basis and ab-initio pseudopotentials
\cite{fhi-dft}: this technique
should allow for a correct description of solid
aggregates of G's, resembling those of long-range 
deoxyguanosine fibers, provided the single molecules
and the clusters are well described.
Before simulating the structures of our interest, which
are planar ribbons and stacked sequences of G's, we check that 
our technique is well suitable to
describe isolated G molecules as well 
as pairs of molecules in different 
(lateral or vertical) configurations. We then consider 
(GG)$_n$ vertical stacks,
and (GG)$_n$ isolated and stacked periodic ribbons. 
We study the stability of different
vertical configurations as a function of the relative atomic positions 
between adjacent G molecules,
and identify the role of $\pi$-$\pi$ coupling.
We demonstrate that pseudopotential plane-wave 
Density Functional Theory (DFT) calculations at
the current level of accuracy are able to
reproduce not only the equilibrium 
length of isolated hydrogen bonds,
but also periodic sequences of such bonds.
Concerning the electronic properties, all the model solids that 
we considered are semiconducting with large energy gaps. 
The dispersion of the highest valence and lowest conduction bands is 
always negligible in the $(x,y)$ plane containing 
the molecules (the guanines are connected laterally by H bonds). 
Conversely, the dispersion along the $z$ direction perpendicular 
to the G planes found to be extremely 
sensitive to the detailed geometric stacking of the bases, which in turn can 
be affected by different environments. 
In vertical stacking geometries that are similar to 
those present in DNA, the calculated dispersion of both the 
valence band maximum (VBM) and conduction band minimum (CBM) 
are extremely small. Consequently, both electrons and holes introduced by 
``doping'' or photoexcitation would have very large effective masses 
-- hence low mobility -- in these types of structures. 
Instead, our results show that a band-like contribution 
may be responsible for the conduction mechanism in the 2'-deoxyguanosine 
lipophilic derivative \cite{prl}. 

The paper is organized as follows. Section \ref{sec-method} describes 
the computational method. 
Section \ref{sec-results} deals with the calculated 
equilibrium geometries and the electronic structure for the isolated 
G molecule and for the model guanine assemblies. 
Section \ref{sec-discussion} presents a discussion 
of our results with an outlook to 
the implications for the physics of guanine 
ribbons and DNA-like stackings. 
Finally, section \ref{sec-conclusions} contains a summary of the arguments 
presented in the paper. 

\section{Method.}
\protect\label{sec-method}
Our calculations are based on the DFT 
in the Local Density Approximation (LDA) \cite{dg-dft}. 
For the hydrogen-bonded pairs and ribbons, we take into 
account BLYP gradient corrections to the 
exchange-correlation functional \cite{gga-blyp}. 
The electron-ion interaction is described via ab-initio 
norm conserving pseudopotentials in the 
factorized form of Kleinman and Bylander \cite{pseudo}. 
The search for optimized metastable structures is performed 
by total energy minimizations with respect to the ionic and electronic 
degrees of freedom. The former are represented by the 
ionic coordinates, the latter by the 
electronic wavefunctions. The ions are treated 
in a classical formalism, and are displaced 
according to the forces derived from the potential 
determined by the full quantum mechanical electronic structure 
\cite{fhi-dft}, within a Car-Parrinello-like scheme \cite{cp-scheme}. 
For any selected geometry, all the atoms are allowed 
to relax, until the forces 
vanish within an accuracy of 0.05 eV/\AA. 
Thus, for each metastable structure, we obtain 
both the geometry and the consistent single-particle 
electron energies and wavefunctions. 
The electron wavefunctions are expanded in a basis of plane 
waves with kinetic energy up to 50 Ry. This cutoff 
is very high with respect to standard first-principle calculations 
for most solids, and is due to the presence of the first-row elements 
C, N, and O, whose valence electron wavefunctions 
have strong oscillations in the region around the nucleus, thus 
needing many plane waves for an accurate treatment. 
For one of the model systems, 
we have verified that increasing the precision of the calculations both in the 
plane-wave kinetic energy cutoff (up to 60 Ry) and in the accuracy within which the 
atomic forces vanish (0.025 eV/\AA), gives changes of the bond lengths smaller 
than 0.01 \AA~and of the bond angles smaller 
than 0.5 degrees, within a guanine molecule, whereas 
the inter-planar distance in the stacks is not affected at all. 
Additionally, for the 
diatomic molecules N$_2$ and O$_2$ the employed pseudopotential was tested 
up to a cutoff of 80 Ry, with no significant improvement. 

Our method employs periodic boundary conditions in 
three dimensions. In order to simulate 
isolated molecules, we choose a large supercell 
of size 15.9 \AA~$\times$ 15.9 \AA~$\times$ 10.6 \AA. 
Such a choice ensures that the minimum distance (in any spatial 
direction) between two molecules is larger than 8.5 \AA. In particular, 
the distance is 10.6 \AA~ in the direction perpendicular 
to the plane of the molecules, very large with respect to the 
distance of 3.37 \AA~ between two neighboring bases in B-DNA. 
The supercells for different assemblies of G molecules are 
described in section \ref{sec-results}, where the 
model structures are discussed in detail. 
For Brillouin Zone sums, the single high-symmetry $\Gamma$ point 
has been employed in the case of isolated molecules 
and dimers, while one or two (depending on the symmetries) special 
{\bf k} points \cite{monk-pack} have been 
employed in the case of periodic columns and ribbons, and in 
the case of stacked ribbons. 

The computational technique has been successfully applied 
in many investigations of the structure and electronic 
properties of inorganic and organic materials 
\cite{dg-dft,review-dft-inorganic,review-dft-molec}. 
In section \ref{sec-results}, before reporting the results 
of our calculations for the model guanine assemblies, we 
address the issue of extending such calculations to biomolecules, 
by presenting a test on the G base. 
%
\section{Results.}
\protect\label{sec-results}
In this section we present our results about the 
isolated G molecules and about their assemblies 
and discuss them in the frame of 
the existing theoretical literature, 
which is limited to isolated G's 
and G-pairs \cite{sponer2,sponer3,sugiyama-saito} 
and does not give any account of periodic G columns and ribbons. 
We relate the outcome of the calculations 
to new experimental findings about the 
structure and electrical behavior 
of deoxyguanosine-based solids \cite{prl}. 

The section is divided into sub-sections for the different systems: 
(A) the isolated G molecule; (B) the stacked GG dimers; 
(C) the stacked (GG)$_n$ columns; (D) the hydrogen-bonded GG pairs; 
(E) the hydrogen-bonded (GG)$_n$ ribbons and the stacked ribbons. 
\subsection{The Isolated Guanine Molecule.}
\protect\label{subsec_1mol}
We start by presenting the calculated structure 
and selected electron states for the G molecule. 
This simple system allows us to evaluate the 
accuracy of our method, by comparison with 
X-ray data and with the 
outcome of previous ab-initio calculations. 
Furthermore, the total energy of the equilibrium 
structure of the isolated molecules is necessary 
in order to evaluate formation energies of 
dimers, ribbons, and columns. 

The structure of guanine is well known from 
X-ray studies \cite{saenger,guaninexp}. 
Figure \ref{1mol.f}(a,b) shows a plane view of the molecule, and the 
isosurface plot for the total charge density. 
The calculated bond lengths and angles are 
in good agreement with X-ray data (see Table \ref{Gmol.t}). Most of 
the bond lengths are {\sl underestimated} within 2\% and 3\% 
with respect to the experimental structure, 
the only exceptions being 
the {\sl underestimate} 
of 4.3\% for the C$_6$-O$_6$ bond and the {\sl overestimate} of 1.6\% 
for the C$_6$-N$_1$ bond. 
The average C-H and N-H distances, not reported in Table \ref{Gmol.t}, 
are 1.08 \AA~ and 1.01 \AA, respectively, in good agreement 
with bond lengths in NH$_3$ and CH$_3$ molecules. 
The bond angles in guanine are also 
reproduced with a high degree of accuracy (Table \ref{Gmol.t}): 
the discrepancy with respect to the experimental 
data is below 2.5\%, but the average percentage 
error is 1\%. 
The formation energy of G, with respect to its 
elemental components in stable phases (O$_2$, H$_2$, and N$_2$ molecules, 
crystalline diamond) is 87.5 Kcal/mole. 
By starting the atomic relaxation with a planar G molecule 
as the initial condition, we find that the planar 
configuration is indeed a metastable state. By considering 
a different initial condition with a non-planar amino group 
(the NH$_2$ complex bonded to the site C$_2$, see Figure \ref{1mol.f}a), 
we find another metastable state. However, the 
deviation from planarity is small, and the total 
energy difference between the planar and the puckered 
geometries is also very 
small, within the precision of our calculations (estimated to be 
about 10 meV/G \cite{note-eform}). While previous calculations have pointed 
out a stronger stabilization effect of non-planarity 
\cite{sponer3}, no direct gas phase experimental data 
exist. Moreover, the hydrogen bonds tend to flatten the structure 
when forming dimers and ribbons of G's; thus, we are confident 
that our results for the periodic structures 
are not affected by this issue. 

The single particle eigenvalues are characterized by 
a DFT-LDA energy gap of 4.8 eV between the 
highest occupied (HOMO) and the lowest unoccupied (LUMO) 
electron states \cite{dg-dft}. 
We have identified $\sigma$ and $\pi$ orbitals. 
The HOMO (see Figure \ref{1mol.f}c) has 
a $\pi$ character and is localized 
on the C$_8$, O$_6$, N$_2$ atoms, and on the C$_4$-C$_5$ 
and N$_3$-C$_2$ bonds. The LUMO (see Figure \ref{1mol.f}d) 
has a $\pi$ character and 
is localized on the N$_9$, C$_4$, C$_5$, C$_2$, and N$_1$ atoms. 
Because they extend out of the guanine plane, 
both the HOMO and the LUMO states are well suitable for 
interactions with adjacent similar states when forming vertical stacks, 
inducing a splitting of degenerate molecular orbitals by an amount 
that depends on the strength of such interactions. In the case of an 
infinite periodic stack, this is the mechanism 
that may give rise to band dispersion for sufficiently strong coupling, 
and to mobile carriers if the bands are partially occupied 
(for instance, as a consequence of doping or photoexcitation). 
\subsection{Stacked GG dimers.} 
\protect\label{subsec_GGstackedimers}
In order to select low-energy geometries 
for the vertical columnar structures, whose electronic properties 
are the main subject under investigation, 
we have first considered stacked dimers. 
We name stacked dimer a pair 
of G molecules lying in parallel planes whose distance 
in the perpendicular direction is an output of our calculations. 
A column is obtained by periodically replicating a dimer 
along the stacking (perpendicular) direction. 

We have analyzed several configurations 
(Figure \ref{geometries.f}) characterized 
by the relative azimuthal rotation angle of 
the two G's in the pair, with respect to 
an axis perpendicular to the G plane, and by an in-plane translation. 
The supercell used in the calculations was 
15.9 \AA~$\times$ 15.9 \AA~$\times$ 19.1 \AA. With 
periodic boundary conditions, two dimers in 
neighboring supercells are 15.7 \AA~apart in the 
stacking direction, sufficient to avoid 
spurious interactions between them. 
By allowing all the atoms to relax, 
the interplanar distance between 
the two G's in a pair was a free parameter. 
In this way we determined the equilibrium interplanar distance 
to be used in the calculations for the periodic columns. 
For all the configurations shown in Figure \ref{geometries.f} 
({\sl e.g.}, independently of the relative rotation angle 
between the bases), 
the average interplanar distance was 3.37 \AA, typical 
of base stacks in B-DNA. 
Each base maintained a planar geometry, 
with out-of-plane fluctuations smaller than 0.05 \AA. 

Top views of the computed stacked dimers 
are illustrated in Figure \ref{geometries.f}. 
In Figure \ref{geometries.f}a, the two G's 
of the dimer are perfectly eclipsed, 
in particular the hexagonal rings are on top of each other, and there is 
maximum superposition of the $\pi$-like HOMO and LUMO 
of the two guanines (label GGv.A).  
In Figure \ref{geometries.f}b, the rotation angle 
is zero as for GGv.A, but the center of mass of one molecule
is shifted with respect 
to the other, in order to lower the $\pi$-$\pi$ superposition 
(label GGv.B). 
In Figure \ref{geometries.f}c, the azimuthal rotation angle is zero, as 
well as the relative translation of the two G's, but 
there is a reflection of the upper molecule with respect to 
its plane: this configuration (label GGv.C) 
is the only one, among those considered in this work, 
that exhibits a reflection and allows to discriminate between 
two molecular faces. 
In Figure \ref{geometries.f}d, the rotation angle 
is 180$^{\circ}$, and the $\pi$-$\pi$ superposition is 
large, though smaller than in GGv.A (label GGv.D). 
In Figure \ref{geometries.f}e, the azimuthal rotation angle 
is 36$^{\circ}$, and a translation 
brings the two molecules in a 
configuration similar to that in B-DNA (label GGv.E). 

Although some of these stacked structures are not likely to occur 
in nature, they allow to understand important 
microscopic features that may be relevant in real structures. Most 
notably, the dependence of the stability and of the electronic properties 
on the azimuthal angle and $\pi$ superposition is accessible. 
The lowest-energy configuration among 
those shown in Figure \ref{geometries.f} is GGv.D, 
while GGv.A has the highest formation energy, and 
the other dimers have intermediate formation energies \cite{note-faces}. 
However, the energy difference 
between the two extreme cases GGv.A and GGv.D is small, 
about 250 meV/G. 
We attribute the highest formation energy of GGv.A 
to the electrostatic repulsion due to the 
complete $\pi$-$\pi$ superposition. 
In fact, the $\pi$-like HOMO's of guanines in 
neighboring planes are mostly responsible for their interaction: 
the superposition of negative charge in the same region 
of space (the hexagonal ring) for configuration GGv.A contributes with 
a Coulomb repulsion. 
The $\pi$-$\pi$ 
superposition is large also in GGv.D, but the 
repulsion is much smaller, making GGv.D a more viable 
model for a stacked GG pair \cite{sponer2}. 

We wish to point out that, although the lower 
superposition of adjacent $\pi$ orbitals of GGv.D with respect 
to GGv.A decreases the electrostatic repulsion between 
the two molecules and makes the dimer more viable, such 
interaction is still rather large and is the origin 
of energy dispersion in columnar structures based on the GGv.D dimer. 
This issue is addressed 
in the next sub-section, by computing the band structures 
of model periodic stacks built up with the dimers just described. 
It is also worth stressing that, for the stacked dimers, 
without inclusion of sugars, phosphates, 
and water as in real situations, the rotation angle of 36$^{\circ}$ 
is not preferred. 
This result is in line with the fact that the inner core 
of the base pair stack is not completeley responsible 
for the stability of double-stranded DNA, but also the backbone 
and the environment are relevant factors. 
\subsection{Periodic stacked poly(G) columns.} 
\protect\label{subsec_GGstackedcolumns}
The building block of each periodic stack is a GG pair; 
we have analyzed five different model columns 
starting from the dimers described above (Figure \ref{geometries.f}). 
The supercell 
for these calculations is 15.9 \AA~$\times$ 15.9 \AA~$\times$ 6.74 \AA: 
no vacuum region separates two adjacent dimers in the stacking direction. 
The periodic structures are labeled with the same 
names as the stacked dimers. Direct and reciprocal one-dimensional 
crystal lattices are associated to the periodic columns: the 
basis vectors of these lattices are {\bf a}$_3$=6.74 \AA, and 
{\bf b}$_3$=$\frac{2\pi}{a_3}\hat{\bf \rm a}_3$.

The formation energies are reported in Table \ref{energetics.t}. 
The energetical order is the same as for the dimers. 
The most stable configuration is GGv.D. 
We stress again that configuration A, with 
very large $\pi$-$\pi$ superposition, has the highest 
formation energy. 
The energy is substantially reduced in the column GGv.D, 
where the superposition is still large but the 
repulsion is weaker: 
the hexagonal rings lie on top 
of each other, but the O atoms in adjacent planes lie 
opposite to each other. Thus, we attribute the 
dominant repulsive contribution 
of the $\pi$-$\pi$ interaction to the O atoms. 

A detailed analysis of the electronic properties 
reveals other interesting features. 
We report the numerical data in Table \ref{energetics.t} 
and we show the bandstructure for columns GGv.A, GGv.D, and GGv.E 
in Figure \ref{bandstructure.f}. 

GGv.E, with the two G's rotated by 36$^{\circ}$ as in 
portions of nucleic acids, 
has flat HOMO- and LUMO-derived bands, very large 
effective masses, and is therefore incompatible 
with even a partial band transport mechanism for 
the conduction in guanine-based devices \cite{prl}. 
The periodic columns GGv.A and GGv.D, instead, 
have electronic properties that may support 
electronic transport through the base stack. 
For GGv.A, the HOMO-band disperses downwards 
by 0.65 eV and the LUMO-band disperses upwards 
by 0.52 eV, between the center ($\Gamma$) and 
the edge (A=$\frac{1}{2}${\bf b}$_3$, {\bf b}$_3$ being 
the reciprocal lattice vector along the stacking 
direction) of the Brillouin Zone. For GGv.D, the dispersions 
of the HOMO- and LUMO-bands are, respectively, 0.26 eV 
downwards and 0.13 eV upwards. 
The effective 
masses of GGv.A and GGv.D, reported in Table \ref{energetics.t}, 
are of the order of 1$\div$2 free electron masses. 
Although these values are much larger than those of conducting 
organic polymers \cite{polymers}, they are similar to 
those of inorganic materials for which band transport 
is demonstrated. 
Notable examples of such materials are wide-bandgap 
semiconductors, such as group-III nitrides \cite{nitrides}. 
As a note to support the strength of our results, we wish to point out 
that a test calculation for the model GGv.A demonstrates that the 
values of the band dispersions do not depend (within 0.02 eV) on the 
accuracy required for vanishing atomic forces (0.025 eV/\AA instead of 0.05 eV/\AA), 
and to the approximation employed for the exchange-correlation functional 
(BLYP instead of LDA). 

To illustrate how 
the $\pi$-$\pi$ stack may originate channels for 
charge migration through a band transport mechanism, 
in Figure \ref{HOMO_D.f} we show an isosurface of the HOMO state at 
the A point 
for the columnar structure GGv.D: 
the interaction between the two molecules 
in the cell is evident in the superposition 
resulting in a delocalized orbital. 

Summarizing our study of the columnar structures, we 
emphasize that the orbital interaction 
is strong. At variance, we 
will soon show that it is practically absent 
in hydrogen-bonded pairs. This is in agreement 
with the common knowledge that hydrogen bonds 
have an electrostatic (rather than covalent) nature.
In the following subsections, we compare 
the behavior of planar G pairs and ribbons with that of stacked 
G pairs and columns. 
In particular, we find 
that hydrogen bonding does not give origin to 
dispersive electron bands, contrary to 
what we have seen for the $\pi$ stacking. 

\subsection{Planar hydrogen-bonded GG dimers.} 
\protect\label{subsec_GGplanardimers}
We have selected one possible arrangement 
of hydrogen bonds, giving a structure named 
GG3 \cite{sponer1,machado} (Figure \ref{dimerplanar.f}a). 
We have not considered other documented 
hydrogen-bonded GG pairs \cite{sponer1}, 
because we focus our interest here in the 
guanine ribbons present in the fibers of 
a lipophilic derivative
of 2'-deoxyguanosine: Such fibers \cite{gottarelli00}
have the bond network illustrated 
in Figure \ref{dimerplanar.f}(b). 
The equilibrium structure that we 
obtain after atomic relaxation is in good agreement with that 
of the GG3 hydrogen-bonded pair previously 
described via quantum chemistry and DFT 
cluster calculations \cite{sponer1}. 
The two individual G molecules 
in the planar pair 
remain very similar to their isolated form. 
The N7(H)$\cdot\cdot\cdot$N1 hydrogen bond has 
a length of 2.93 \AA~and forms an angle of 177$^{\circ}$ 
(2.96 \AA~and 173$^{\circ}$ in the theoretical literature \cite{sponer1}). 
The N2(H)$\cdot\cdot\cdot$O6 hydrogen bond 
has a length of 2.85 \AA~and forms an angle of 169$^{\circ}$ 
(3.27 \AA~and 166$^{\circ}$ in the 
theoretical literature \cite{sponer1}). 
Our calculated value for the DFT-BLYP energy gap \cite{dg-dft} of the GG3 
dimer is 2.45 eV. 
The formation energy of the structure, with respect to 
two isolated guanine molecules, is -310 meV/G. 
This value, compared to the formation energy 
of the most stable stacked dimer GGv.D 
($\simeq$ -200 meV/G), is in agreement 
with the previous demonstration that hydrogen bonds are 
stronger than stacking interactions \cite{sponer1}. 

The structure of the GG3 dimer was calculated 
only as a basic building block of the fiber-state 
ribbons, whose electronic properties we are interested in. 
Thus, we believe that the underestimate of the N2(H)$\cdot\cdot\cdot$O6 
distance with respect to the results of other computations 
is not a serious issue. A slight underestimate 
of the hydrogen-bonds calculated in the frame of DFT 
is documented \cite{sponer1}. Furthermore, 
the contraction of the N2(H)$\cdot\cdot\cdot$O6 hydrogen bridge 
is due to an imprecise account of secondary interactions 
such as C8(H)$\cdot\cdot\cdot$O6. 
It is likely that such 
secondary interaction is reduced in 
one-dimensional ribbons, thus minimizing the shortcomings of DFT. 

\subsection{Planar hydrogen-bonded ribbons.} 
\protect\label{subsec_GGribbons}
In a periodic ribbon obtained by piling up replicas 
of the GG3 dimer (Figure \ref{dimerplanar.f}b), 
the equilibrium structure 
maintains the bonding characteristics of the dimer. 
There is a huge energy gain in forming the one-dimensional 
ribbon, of 820 meV/G with respect 
to isolated G molecules, and of 510 meV/G 
with respect to the hydrogen-bonded GG3 pair (Table \ref{nastri_ene.t}). 
Note that G has a large dipole moment \cite{sponer3}; 
the dipole moments of G's add up to give a non-vanishing 
dipole also in the GG3 dimer and in the ribbon, parallel 
to the ribbon axis \cite{prl}. Such electrostatic interactions 
account for the high stability of the planar hydrogen-bonded structures, 
well known both in solution and in the solid state. 

The bandstructure of the ribbon was calculated along a symmetry line 
parallel to its axis. 
The HOMO- and LUMO-derived bands are separated by a DFT-BLYP energy gap 
\cite{dg-dft} of 
3.84 eV, and are both dispersionless. Consequently, 
these bands 
have practically infinite effective masses, so that the electrons and holes 
in these states are not mobile according to this picture. 
The electronic state analysis shows that the 
HOMO and the LUMO have a $\pi$ character,
similar to isolated G. The electronic states are localized 
around single G molecules,
no delocalized intermolecular states, 
extended through the hydrogen bonds in the ribbon, are present. 
The dispersion induced by hydrogen-bonding is not compatible with 
band transport. The only possible conduction mechanism 
through hydrogen bonds is via electrostatic interactions. 
Therefore, in a device where dried deoxyguanosine fibers 
are deposited between two metal electrodes \cite{prl}, if the ribbons 
are stretched between the electrodes, band transport 
cannot contribute to conduction. 

In order to investigate the competition of
$\pi$-$\pi$ coupling versus hydrogen-bonding interactions, 
we have simulated two 
different configurations of stacked ribbons, periodic 
along the stacking direction. Top views are shown 
in Figure \ref{dimerplanar.f}(b,c). 
Figure \ref{dimerplanar.f}b shows indeed a single 
ribbon, but the top view is equivalent for two 
exactly eclipsed ribbons on top of each other (label SR.A): 
the stacking between adjacent bases is similar to 
that of column GGv.A, and all the individual bases are perfectly aligned. 
Figure \ref{dimerplanar.f}c shows a stack of two ribbons 
where only half of the individual G molecules 
lie on top of each other, 
with a stacking similar to that of configuration GGv.D 
(this structure is labeled SR.D). In both the SR.A and SR.D 
geometries, the projections of the ribbon axes on the 
$(x,y)$ plane coincide, identifying an axis for the stack: 
this axis defines the $\Gamma-X$ direction in the 
one dimensional BZ. 

The supercells are 21.2 \AA~$\times$ 11.3 \AA~$\times$ 6.74 \AA~ for SR.A
and 24.3 \AA~$\times$ 11.3 \AA~$\times$ 6.74 \AA~ for SR.D. 
The choice of different supercells allows to have 
a vacuum region of the same volume between 
neighboring ribbons in the two configurations SR.A and SR.D. 
The relaxed structures maintain the geometry of the single ribbon:
no G deformations, no variations of H-bond lengths and angles, 
and no out-of-plane buckling are observed. This 
finding indicates that the stacking does not affect 
the hydrogen-bonding mechanism that determines the 
structure of the isolated ribbon. 

Numerical data for the energetics and the electronic properties 
are reported in Table \ref{nastri_ene.t}. 
Both structures SR.A and SR.D are energetically favorable 
with respect to isolated G molecules and to stacked columns, 
but unfavorable with respect to isolated ribbons, 
as indicated by the smaller energy gain of the stacked ribbons. 
The configuration SR.D is more stable than SR.A by 280 meV: 
this trend is consistent with what found previously for the 
columnar stacks, where a higher stability 
is achieved by lowering the $\pi$-$\pi$ superposition 
in building the stacks. Although the configurations SR.A and SR.D 
that we have calculated here are not viable models for piling up 
ribbons into solid-state crystals, ordered phases for 
deoxyguanosine fibers are known to exist \cite{gottarelli98}. 
We note that other factors, such as presence of the phosphate-sugar 
backbone, or the possibility of different relative positions 
of the neighboring ribbons, should be taken into account 
to achieve a full description, which is beyond the scope of 
this work. 

In Figure \ref{2nastri_bande.f} we show the calculated
bandstructure for the SR.A and SR.D configurations. 
In both cases, the HOMO- and LUMO-derived bands are 
dispersionless along the $\Gamma-X$ direction, 
and the charge carriers are not mobile in this direction. 
This behavior is the same as in isolated ribbons: 
therefore, it is a further evidence that 
the base-base interactions
due to the stacking do not change the features 
of the ribbons. 
In the $\Gamma-A$ direction, parallel to the stacking direction, 
the HOMO- and LUMO-derived bands are dispersive, as in the columnar 
structures of subsection \ref{subsec_GGstackedcolumns}. 

The bands for structure SR.A in the 
$\Gamma-A$ direction are more dispersive than those for structure 
SR.D, due to the higher $\pi$-$\pi$ superposition. 
As a consequence, the corresponding electron and hole 
effective masses along the $\Gamma-A$ direction are smaller. 
The effective masses of the stacked-ribbon periodic structures 
are very similar to those of the stacked periodic columns 
(subsection \ref{subsec_GGstackedcolumns}): this is an 
indication that the hydrogen-bonding network of the individual ribbons 
does not affect the electronic properties in the perpendicular 
direction. The only exception to this trend is m$_e$, 
which becomes infinite for the stacked ribbons SR.D while it is 
finite for the equivalent stacked column GGv.D: 
this is due to the fact that in structure SR.D one half 
of the individual bases are on top of each other 
(see Figure \ref{dimerplanar.f}c), 
while the other half do not participate in the $\pi$-$\pi$ coupling. 

Summarizing the results discussed above, we wish to 
point out that the $\pi$-$\pi$ coupling and H-bonding 
interaction mechanisms are not in competition. 
H-bonding has an electrostatic nature and accounts 
for in-plane stabilization: it is not compatible 
with a band transport mechanism, 
and it is not modified by base stacking. 
$\pi$-$\pi$ coupling is weaker than H-bonding: 
it is compatible with band transport along the stacking 
direction and it is not affected by the local 
details of the planar base sequence. 

\section{Discussion.} 
\protect\label{sec-discussion}
\subsection{Guanosine films.} 
\protect\label{subsec-guanosine}
As outlined in the Introduction, the recent lively research 
activities in molecular electronics have pointed out peculiar guanine 
assemblies to be exploited as electrical conductors. 
In fact, 
the experiments on 
a lipophilic derivative
of 2'-deoxyguanosine \cite{prl}, demonstrated 
conduction through the biomolecular material 
deposited in a nanogate between metal electrodes. 
The details of the conductivity depend on the 
on the experimental conditions, in particular on the gate width. 
Our results indicate that the observed conductivity, 
resembling that of a semiconductor in the intermediate 
gate length regime (few hundreds nm) and that of a diode junction 
in the short gate length regime (less than 100 nm), 
is compatible with a Bloch contribution to the 
transport mechanism. 
We have shown that 
delocalized orbitals through the base stack may be formed, 
provided that 
a relevant superposition of 
the hexagonal rings of the single guanines is maintained. 
Therefore, if the guanosine ribbons in the device gate align locally 
with their plane perpendicular to the direction connecting the electrodes, 
in a geometry similar to that proposed as SR.A, then extended 
Bloch-like states may be partially responsible for the observed 
conduction. This stacking orientation does not need to be complete 
in order for the proposed mechanism to work: it is sufficient 
that randomly aligned stacks form locally, and that the 
total resulting component in the direction connecting the electrodes 
is non vanishing. 
In the gate length regime ranging from 100 nm to 300 nm, 
semiconducting-like conductivity is revealed: 
we propose that, in such a condition,  
the ribbons are locally stacked in such a way to form 
partially delocalized orbitals that give a global band-like 
contribution. 
In the gate length regime below a 100 nm, a diode-like behavior 
is observed: 
this characteristic is assigned 
to an interaction of the $\pi$ 
stack with the total dipole moment of the ribbons, as 
discussed elsewhere \cite{prl}. 

\subsection{DNA molecules.}
\protect\label{subsec-DNA}
While our study is limited to a single type of nucleoside, 
it also allows a discussion of the electronic properties 
of DNA molecules, where G sequences in base stacks play an important role 
\cite{grinstaff-rev99,barbara-rev99,schuster-polaron}. 
Depending on the energetics of the base sequence, and on the overall 
structural aspects of the system under investigation, 
the mechanisms proposed for DNA-mediated 
charge migration 
include single-step superexchange \cite{lewis,marcus}, 
multistep hole hopping \cite{jortner98,bixon-hhop99}, 
phonon-assisted polaron hopping \cite{schuster-polaron}, and 
band transport 
\cite{porath-nat00}. 

In DNA double strands, the relative arrangement of neighboring bases 
along the axis of the helix \cite{saenger} is characterized by a 
rotation angle around 36 degrees. Our results show that 
periodic G columns in such a configuration do not 
support the formation of extended molecular orbitals. Therefore, 
band transport would not be effective. For what concerns 
the mechanism of conductivity in DNA molecules, our results support 
the conclusion that the contribution of band transport would be 
very small \cite{ye,bixon-hhop99,dePablo}, in contrast to 
recently proposed interpretations of experimental data 
\cite{porath-nat00}, unless structural distortions, possibly 
activated by temperature effects, may induce rotations that 
support the formation of partially delocalized electron states. 
Although in real DNA molecules the backbone 
phosphate and sugar groups may affect the overall 
charge mobility, we believe that the presence of the outer mantle would not 
change our present conclusion that band transport 
is not supported by the native DNA stack: eventually, it may 
contribute a hopping or ionic mechanism, but it is unlikely 
to contribute to the formation of extended electron orbitals. 
The importance of the base stack for electron transfer 
through DNA molecules has been highlighted in recent 
theoretical invesigations 
\cite{ye}. 

\section{Conclusions.} 
\protect\label{sec-conclusions}
We have reported the results of ab-initio calculations for the structure, 
energetics, and electronic properties of several 
guanine assemblies. 
The geometry of isolated molecules and hydrogen-bonded 
dimers is well described with our technique, based 
on plane-wave pseudopotential density functional theory, 
as demonstrated by comparison to available 
theoretical and experimental data. 

We found that hydrogen-bonding and 
$\pi$-$\pi$ coupling are independent 
mechanisms that control the self-assemblying of guanine bases. 
Hydrogen-bonding is not 
responsible for band transport: In fact, we have 
shown that no band dispersion is 
present along a planar ribbon of H-bonded guanines. 
Instead, base stacking is accompanied by 
$\pi$-$\pi$ interactions that, 
for the case of sufficiently large overlap between adjacent $\pi$ orbitals 
({\sl e.g.}, GGv.D configuration), induce energy dispersion and are 
consistent with charge mobility. 
Therefore, band transport may be partially responsible 
for charge mobility in nucleotide aggregates, in structures 
characterized by a large base-base superposition. This mechanism 
is likely complemented by hopping to connect (through space) different 
regions where such superposition is realized. 

\section{Acknowledgements.}
\protect\label{sec-ack}
We acknowledge the allocation of computer resources
from INFM Progetto Calcolo Parallelo. 
Fruitful discussions 
with M. Buongiorno Nardelli, R. Cingolani, 
G. Gottarelli, and R. Rinaldi, are 
also sincerely acknowledged. We are grateful to F. Grepioni 
for communicating results prior to publication. 


%
\begin{table}
\caption{Structural data for the isolated guanine molecule. 
The experimental structure taken as reference \protect\cite{saenger,guaninexp} 
is that for guanosine, with 
a sugar moiety instead of an H atom in position 9 
({\sl e.g.}, attached to atom N$_9$ in Figure \ref{1mol.f}).}
\begin{tabular}{|c|c|c|c|c|c|} 
\multicolumn{3}{|c|} {distance (\AA)} & \multicolumn{3}{c|} 
{angle ($^{\circ}$)} \\ \cline{1-6}
this work & experiment & discrepancy (\%) & 
this work & experiment & discrepancy (\%) \\ \hline
(C$_8$-N$_7$) 1.281 & 1.304 & -1.8 & (N$_9$C$_8$N$_7$) 112.4 & 113.5 & -1.0 \\
(C$_5$-N$_7$) 1.347 & 1.389 & -3.0 & (C$_8$N$_7$C$_5$) 105.5 & 104.2 & +1.2 \\
(C$_5$-C$_6$) 1.419 & 1.415 & +0.3 & (N$_7$C$_5$C$_4$) 110.6 & 110.8 & -0.2 \\
(C$_6$-O$_6$) 1.186 & 1.239 & -4.3 & (N$_7$C$_5$C$_6$) 130.8 & 130.1 & +0.5 \\
(C$_6$-N$_1$) 1.415 & 1.393 & +1.6 & (C$_4$C$_5$C$_6$) 118.6 & 119.1 & -0.4 \\
(C$_2$-N$_1$) 1.345 & 1.375 & -2.2 & (C$_5$C$_6$O$_6$) 131.4 & 128.3 & +2.4 \\
(C$_2$-N$_2$) 1.336 & 1.341 & -0.4 & (C$_5$C$_6$N$_1$) 109.2 & 111.7 & -2.2 \\
(C$_2$-N$_3$) 1.287 & 1.327 & -3.0 & (O$_6$C$_6$N$_1$) 119.4 & 120.0 & -0.5 \\
(C$_4$-N$_3$) 1.323 & 1.355 & -2.4 & (C$_6$N$_1$C$_2$) 127.1 & 124.9 & +1.8 \\
(C$_4$-C$_5$) 1.383 & 1.377 & +0.5 & (N$_1$C$_2$N$_3$) 122.6 & 124.0 & -1.1 \\
(C$_4$-N$_9$) 1.346 & 1.377 & -2.3 & (N$_1$C$_2$N$_2$) 117.0 & 116.3 & +0.6 \\
(C$_8$-N$_9$) 1.361 & 1.274 & -1.0 & (N$_2$C$_2$N$_3$) 120.3 & 119.7 & +0.5 \\
                    &       &      & (C$_2$N$_3$C$_4$) 113.9 & 111.8 & +1.9 \\
                    &       &      & (N$_3$C$_4$C$_5$) 128.5 & 128.4 & +0.1 \\
                    &       &      & (N$_3$C$_4$N$_9$) 126.0 & 126.0 &  0.0 \\
                    &       &      & (C$_5$C$_4$N$_9$) 104.6 & 105.6 & -0.9 \\
                    &       &      & (C$_4$N$_9$C$_8$) 107.0 & 106.0 & +0.9 \\
 \end{tabular}
 \protect\label{Gmol.t}
 \end{table}
\begin{table}
\caption{
Energetical and electronic data of different periodic stacked 
configurations. E$_{form}^1$ is 
calculated with respect to the N$_2$, O$_2$, 
H$_2$ molecules, and C in the diamond phase. 
E$_{form}^2$ is calculated with respect 
to isolated guanine molecules. 
E$_{gap}$ is the energy difference between the HOMO 
and LUMO single particle eigenstates, calculated by DFT-LDA 
\protect\cite{dg-dft}. 
m$_e$ (m$_h$) is the electron (hole) effective mass, in units 
of the free electron mass m$_0$. 
The structures are labeled 
as in Figure \ref{geometries.f}; G is a single molecule.} 
\begin{tabular}{l|c|c|c|c|c|c}
              &  G    & GGv.A & GGv.B & GGv.C & GGv.D &  GGv.E    \\ \hline
E$_{form}^1$ (eV/G)   & -3.80 & -3.69 & -3.79 & -3.93 & -4.07 & -3.89 \\
E$_{form}^2$ (meV/G)  & 0     & +100  & 0     & -130  & -280  & -100  \\
E$_{gap}$ (eV) & 4.8  & 2.97  & 3.71  &  4.12 & 3.63  & 3.54  \\ 
m$_e$ (m$_0$) &       & 1.41  &       &       & 2.80  & $\infty$ \\ 
m$_h$ (m$_0$) &       & 1.04  &       &       & 2.20  & 5.25  \\
\end{tabular}
\protect\label{energetics.t}
\end{table}
\begin{table}
\begin{center}
\caption{Energetical and electronic data of isolated and stacked ribbons,  
as defined in Table \ref{energetics.t}.}
\begin{tabular}{l|c|c|c}
                    & ribbon     &    SR.A      &     SR.D       \\ \hline
E$_{form}^2$ (meV/G) & -820       &   -310       &   -580         \\
E$_{gap}$ (eV)      & 3.84       &   2.61       &   3.12         \\
m$_e$ (m$_0$)       &            &   1.24       &   $\infty$     \\
m$_h$ (m$_0$)       &            &   1.07       &   2.05         \\
\end{tabular}
\protect\label{nastri_ene.t}
\end{center}
\end{table}
\begin{figure}
\caption{(a) Planar view of the isolated guanine molecule, with 
indication of the chemical species. (b) Isosurface plot of 
the total charge density 
from the ab-initio calculation. (c) Isosurface plot of the HOMO. 
(d) Isosurface plot of the LUMO. 
}
\protect\label{1mol.f}
\end{figure}
\begin{figure}
\caption{Top views of different stacked dimer configurations. 
The structures are explicitly defined in the text. 
Gray (black) dots and lines are used to represent 
atoms and bonds in the upper (lower) plane. Columnar structures 
are obtained by replicating the dimer units along the direction 
perpendicular to the plane of the figure. The chemical species 
are read from Figure \ref{1mol.f}(a).}
\protect\label{geometries.f}
\end{figure}
\begin{figure}
\caption{Bandstructure for configurations 
GGv.A, GGv.D, and GGv.E, calculated along the 
symmetry line parallel to the stacking direction. 
Larger dots indicate HOMO and LUMO states. The 
single-particle energies reported in these plots are 
relative to the top of the highest valence band. }
\protect\label{bandstructure.f}
\end{figure}
\begin{figure}
\caption{Isosurface plot of the HOMO state at the point A 
for the columnar structure GGv.D.} 
\protect\label{HOMO_D.f}
\end{figure}
\begin{figure}
\caption{(a) Top view of the hydrogen-bonded GG dimer. 
(b) Top view of the hydrogen-bonded ribbon obtained by periodically 
replicating the dimer, representing also the structure 
SR.A for the stacked ribbons. 
(c) Top view for the stacked ribbon structure SR.D. Black and 
gray dots identify atoms lying in two different parallel planes.} 
\protect\label{dimerplanar.f}
\end{figure}
\begin{figure}
\caption{Bandstructure for configurations SR.A and SR.D,
calculated along the symmetry lines parallel to the 
ribbon direction ($\Gamma-X$)
and to the stacking direction ($\Gamma-A$). Large dots indicate HOMO and
LUMO states.}
\protect\label{2nastri_bande.f}
\end{figure}

\begin{references}
\bibitem{note-counterions} It is worth stressing that, 
in the experiment on deoxyguanosine, neither the polyanionic phosphate 
backbone that connects the nucleoside moieties in DNA, nor the 
necessary counterions, are present. Therefore, conduction 
through the bases is directly evaluated.
\bibitem{note-eform} The unit eV/G that we use here for the 
formation energies means eV per guanine molecule. 
The conversion factor is 1 eV/G = 23.25 Kcal/mole.
\bibitem{note-faces} 
The fact that the dimer GGv.C 
(the only structure in which one molecule is reflected) 
is not energetically favorable with respect to other geometries, 
indicates that the two faces of guanine are inequivalent 
for what concerns the stacking properties.
\end{references}
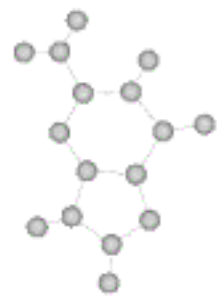
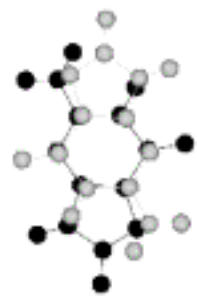
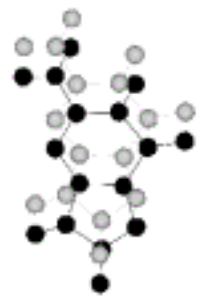
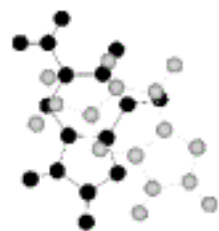
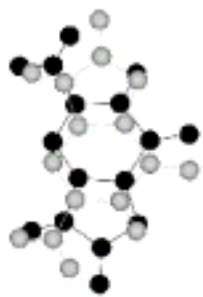
\end{document}